\documentclass[a4paper]{mem}
\usepackage{natbib}
\usepackage{graphicx}
\usepackage[a4paper]{hyperref}
\idline{73}{23}
\begin{document}
   \title{The broad-band properties of the XMM-Newton Bright Serendipitous 
          Sources
\thanks{The XMM-Newton Bright Serendipitous Survey is part of the follow-up program being conducted by the
XMM-Newton Survey Science Center (SSC), an international collaboration
involving a consortium of 10 institutions, appointed by ESA to help the SOC in
developing the software analysis system, to pipeline  process all the
XMM-Newton data, and  to exploit the XMM serendipitous detections,  see
http://xmmssc-www.star.le.ac.uk. The OABrera is one of the Consortium
Institutes.}
}

%\author{R. Della Ceca \inst{1}, 
%           T. Maccacaro  \inst{1},
%	   A. Caccianiga \inst{1},
%	   P. Severgnini \inst{1}, 
%	   V. Braito      \inst{1} , 
%	   on behalf of the XMM-Newton SSC collaboration.
%          M. Marconi \inst{5}\fnmsep
%\thanks{}
%}

\author{   R. Della Ceca, 
           T. Maccacaro,
	   A. Caccianiga,
	   P. Severgnini, 
	   V. Braito, 
	   on behalf of the XMM-Newton SSC collaboration.
%          M. Marconi \inst{5}\fnmsep
%\thanks{}
}

\offprints{R. Della Ceca}
\mail{}

\institute{Osservatorio Astronomico di Brera, via Brera 28, 20121 
Milano, Italy.
\email{rdc, tommaso, caccia, paola, braito: @brera.mi.astro.it}
}

   \abstract{We present here ``The XMM-Newton Bright Serendipitous Survey", 
   a survey program conducted by the XMM-Newton Survey Science Center. 
   In particular we discuss the main goals
of this project, we present the sample(s) and the current optical breakdown
and we discuss some broad band spectral properties  as derived from an hardness
ratio analysis of the single sources. Finally we report  the 4.5--7.5 keV 
Log(N$>$S)-LogS for type 1 and type 2 AGN.
   }
   \authorrunning{R. Della Ceca et al., }
   \titlerunning{The XMM Bright Serendipitous Survey}
   \maketitle
%
%________________________________________________________________

\section{Introduction}

%                                     Two column figure (place early!)
%______________________________________________ Gamma_1 (lg rho, lg e)
%   \begin{figure*}
%   \centering
%   \resizebox{\hsize}{!}{\rotatebox[]{-90}{\includegraphics{figure_hd174005.eps}}}
%   %\includegraphics{empty.eps}
%   %\includegraphics{empty.eps}
%   \caption{dummy image just an example. }
%              \label{FigGam}%
%    \end{figure*}
%______________________________________________________________

Deep {\it Chandra} and {\it XMM--Newton}  observations (Brandt et al., 2001;
Rosati et al., 2002; Moretti et al., 2003; Hasinger et al., 2001; Alexander et
al., 2003) have recently resolved $> \sim 80$\%  of the 2--10 keV  cosmic X-ray
background (CXB) into discrete sources down to  $f_{x}\sim 3\times$10$^{-16}$
erg cm$^{-2}$ s$^{-1}$.
The X--ray data  (stacked spectra and hardness ratios) of these faint samples
are consistent with AGN  being the dominant contributors of the CXB and, as 
inferred by the X--ray colors, a significant fraction  of these sources have
hard, presumably obscured, X--ray spectra,  in agreement with the predictions
of CXB synthesis models  (Setti and Woltjer, 1989; Comastri et al., 2001;
Gilli et al., 2001; Ueda et al., 2003).

However, the majority of the sources found in these medium to deep fields  are
too faint to provide good X--ray spectral information. Furthermore, the
extremely faint magnitudes  of a large number of their optical counterparts
make the spectroscopic identifications very difficult, or even impossible,
with the present day ground--based optical telescopes. Thus, notwithstanding
the remarkable results obtained by reaching very faint X--ray fluxes, the
broad--band physical  properties (e.g. the relationship between optical
absorption and X-ray obscuration and the reason why AGN with similar X-ray
properties have completely different optical  appearance) are not yet
completely understood.  In the medium flux regime ($f_{x}$ between $\sim 10^{-14}$ and 
$\sim 10^{-13}$ erg cm$^{-2}$ s$^{-1}$)  a step forward towards the solution of
some of these problems  has been undertaken  by Manieri et al.,  2002;
Piconcelli et al., 2003, Perola et al., 2004 and Mateos et al., 2004.

With the aim of complementing the results obtained by medium to deep  X-ray surveys, we
have  built the ``The XMM-Newton Bright Serendipitous Source Sample".  We describe below the
main characteristics of this sample and discuss some  of the results obtained so far.
The contribution of this project to the solution of some critical open (and ``hot")
questions like the relationship between optical absorption and X-ray obscuration  and
the physical nature of the ``X-ray bright optically normal galaxies" have been already
discussed in  Caccianiga et al. (2004) and Severgnini et al. (2003),  respectively. 
We stress that
many of these issues are investigated with difficulties using the fainter X-ray samples
because of their typical poor counts statistics for each source.

\section{The XMM-Newton Bright Serendipitous Source Sample}
 
The XMM Bright Serendipitous Source Sample,  a project coordinted by the
{\it Osservatorio Astronomico di Brera}, consists of two flux-limited samples: the
XMM BSS and the XMM HBSS sample having a flux limit of $\sim 7 \times 10^{-14}$
erg cm$^{-2}$ s$^{-1}$ in the 0.5-4.5 keV and 4.5-7.5 keV energy band,
respectively.  This approach was dictated by the need of studying the
composition of the source population as a function of the selection band
and of reducing the strong bias against absorbed sources which occurs when
selecting in soft X-rays.  

Two hundred and thirty-seven  suitable XMM fields (211 for the HBSS) at
$|b|>20$ deg (see Della Ceca et al., 2004) were analyzed and
a sample of 400 sources was selected (see Table 1 for details). 

\begin{table}
\begin{center}
\caption{The current optical breakdown of the BSS and HBSS
Samples}
\begin{tabular}{lrr}
\hline
\hline
                     &           BSS         & HBSS         \\
                     &                       &              \\
                     &                       &              \\
\hline
Objects$^{1}$        &           389         &          67  \\
Area Covered (deg$^2$) &          28.10      &          25.17 \\
\\
Identified:          &           278         &          60  \\
Identification rate  &           71\%        &          90\% \\
                     &                       &              \\
AGN-1                &           180         &          39  \\
AGN-2                &            26         &          16  \\
Galaxies$^{2}$       &             7         &           1  \\
Clusters of Galaxies $^{3}$ &             4         &           1  \\
BL Lacs              &             5         &           1   \\
Stars$^{4}$          &            56         &           2   \\
\hline
\hline
\end{tabular}
\end{center}
$^{1}$  Fifty-six sources are in common between the BSS and HBSS
samples;
$^{2}$  We stress that some of the sources classified as ``Optical Normal
Galaxy" could indeed host an optically elusive AGN (see Severgnini et al.,
2003);
$^{3}$  The source detection algorithm is optimized for point-like 
objects, so the sample of clusters of galaxies is not statistically complete nor 
representative of the cluster population;
$^{4}$  All but one 
of the sources classified as stars are coronal emitters.
\end{table}

The majority of the X-ray sources have enough
statistics (hundreds to  thousands of counts when the data from the three EPIC
detectors are considered) to allow X-ray studies in terms of energy
distributions,  absorption properties, source extent and flux variability.
Moreover  the optical counterpart of $\sim 90$ \%  of the X-ray sources has a
magnitude brighter than the POSS II limit (R $\sim$ 21mag), thus allowing
spectroscopic identification at a 2-4 meter class telescope;  this fact, combined with
the positional accuracy of XMM for bright sources (90\% error circle of 
4$^{\prime\prime}$) implies that, in the large majority of the cases, only one
object  needs to be observed to secure the optical identification.

Up to now 285 X-ray sources have been spectroscopically identified (either from
the literature or from our own observations) leading to a 71\% and 90\%
identification rate for the BSS and HBSS samples respectively (see Table 1 for a 
summary). 
%The optical
%breakdown of the sources identified so far is reported in Table 1

\section{The broad band X-ray spectral properties}

A ``complete" spectral analysis for all the sources in the BSS and HBSS samples
(using data from the two EPIC MOSs and the EPIC pn) is in progress. In the
meantime, and in order to extract first order X-ray spectral information we
report here a ``Hardness Ratio" analysis  of the single sources using only EPIC
MOS2 data. However we note that a ``Hardness Ratio" is often the only  X-ray
spectral information available for the faintest sources in the XMM-Newton
catalogue, and thus, a ``calibration" in the parameter space is needed to
select ``clean" and well-defined samples.

We have used the hardness ratios as defined from the
pipeline processing:
 \[
HR2={C(2-4.5\, {\rm keV})-C(0.5-2\, {\rm keV})\over C(2-4.5\, {\rm
keV})+C(0.5-2\, {\rm keV})}
\]
and
\[
HR3={C(4.5-7.5\, {\rm keV})-C(2-4.5\, {\rm keV})\over C(4.5-7.5\, {\rm
keV})+C(2-4.5\, {\rm keV})}
\]
where C(0.5$-$2\, {\rm keV}), C(2$-$4.5\, {\rm keV}) and
C(4.5$-$7.5\, {\rm keV}) are the ``PSF and vignetting corrected"
count rates in the 0.5$-$2, 2$-$4.5 and 4.5$-$7.5 keV energy bands,
respectively.

Combining the information on HR2 and HR3 (see figure 1) 
we can investigate the broad band 
spectral properties of the sample(s) as well as the selection function(s) of
the BSS and HBSS. From figure 1 it is worth noting that:

$\bullet$ the ``bulk" of the sources optically identified as broad line AGNs (both in
the BSS and HBSS sample) are strongly clustered in a narrow HR2 region between
$-0.75$ and $-0.35$ (corresponding to a photon index in the range between 
$\sim 1.4$ and $\sim 2.5$ if we assume a simple power-law spectral model);  

$\bullet$ all but 2 ($\sim 96$ \%) of the sources classified as stars in the BSS sample
have an HR2 less than $-0.75$. If we assume a simple Raymond-Smith thermal
model, HR2$\leq -$0.75 corresponds to temperatures below $\sim 1.5$ keV, in
very good agreement with the identification as coronal emitters;

$\bullet$ contrary to broad line AGNs and stars, narrow line AGNs seem to be
distributed over a larger area in the HR2-HR3 plane with a well visible
difference in the source position between the Type 2 AGNs in BSS and those in
the HBSS sample (e.g. the large majority of the type 2 AGN in the HBSS have 
HR2 $>-0.3$);

$\bullet$ although many of the narrow line AGN have the hardest spectra
amongst the identified objects, highly suggestive of intrinsic absorption 
(see Caccianiga et al., 2004), there is an indication from this study that a
significant number of narrow line AGN in the BSS sample occupy the locus
typical of X-ray unabsorbed broad line AGN. A deeper investigation of these sources 
is in progress;

$\bullet$ the eleven objects belonging only to the HBSS sample are amongst the 
hardest X-ray sources in the sample. Among these sources there
are 5 Type 2 AGN, one optically normal galaxy,  one BLLac object and 4
unidentified sources.

%
%                                                Two column figure
%----------------------------------------------------------- S_vib   

   \begin{figure*}
   \centering
   \resizebox{\hsize}{!}
   {\includegraphics[clip=true]{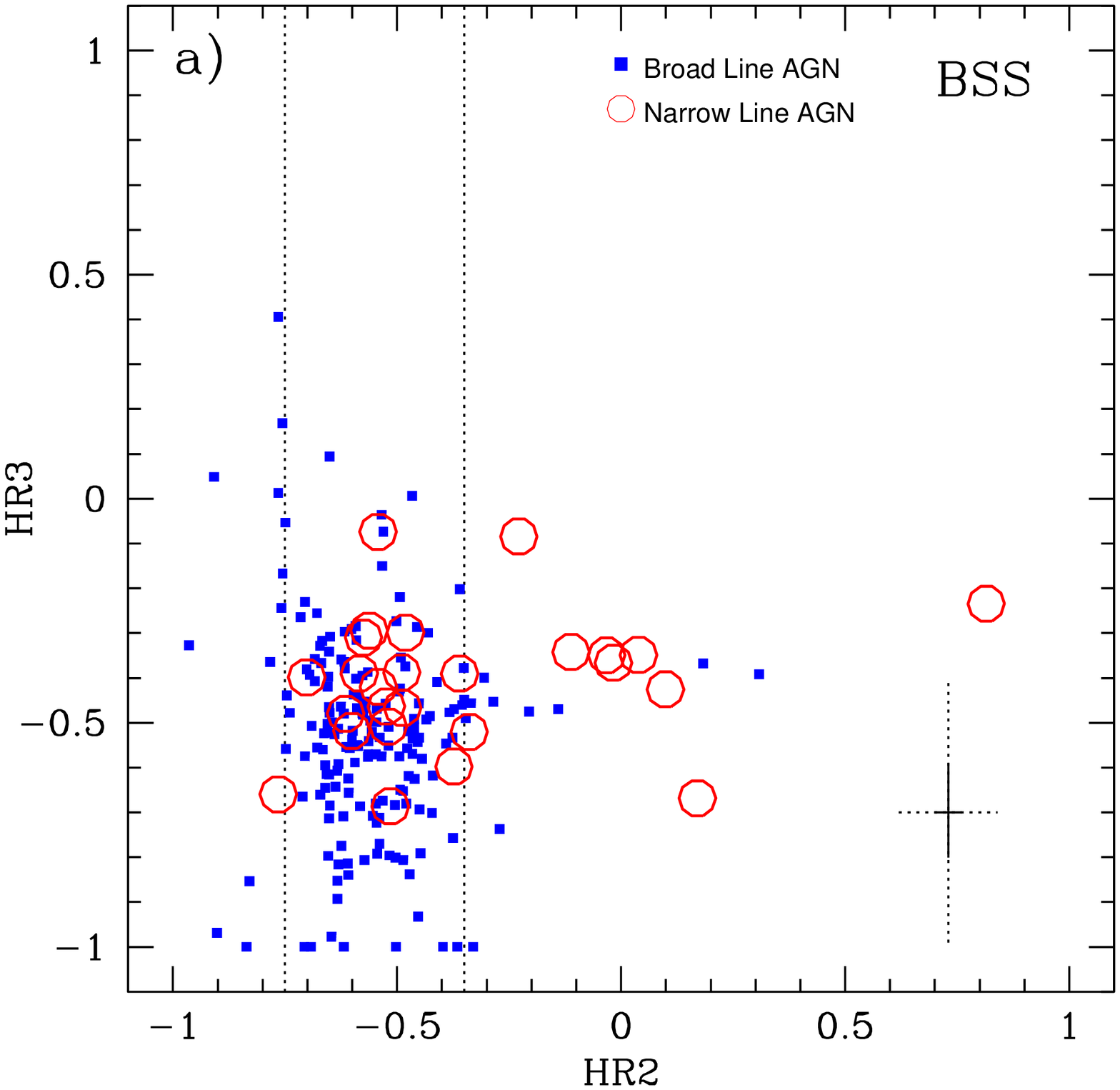}
    \includegraphics[clip=true]{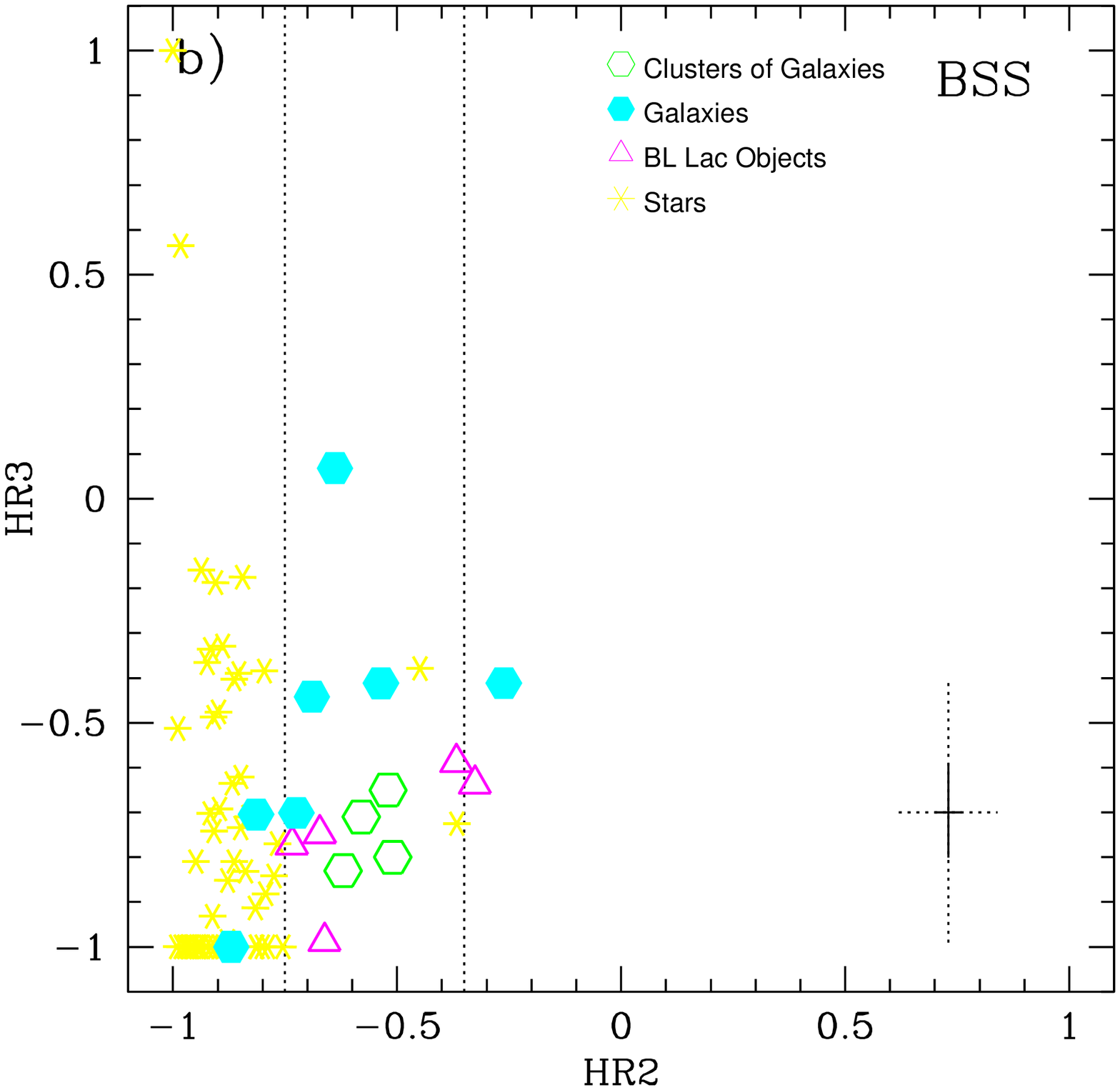}}
    \resizebox{\hsize}{!}
    {\includegraphics[clip=true]{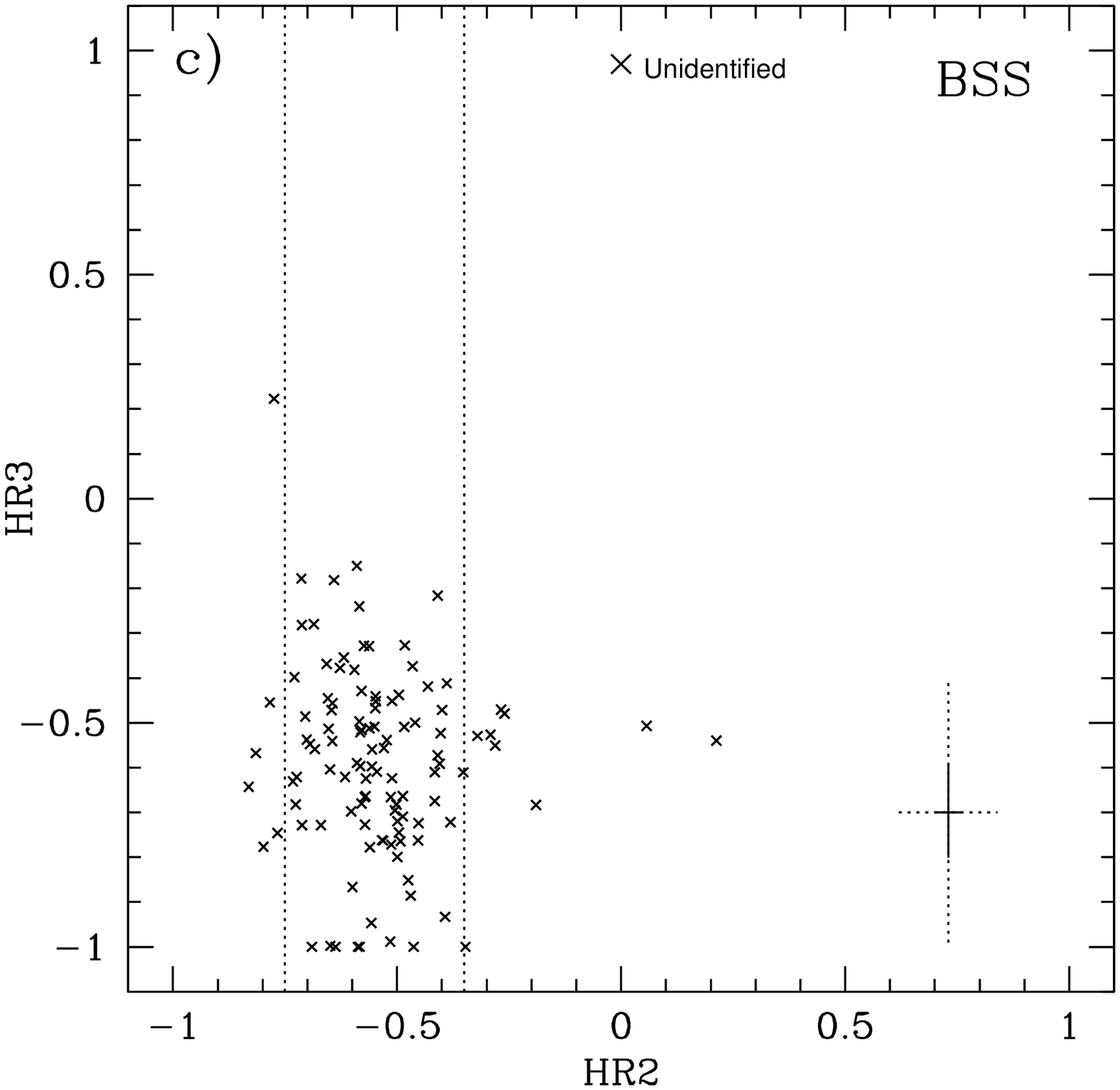}
    \includegraphics[clip=true]{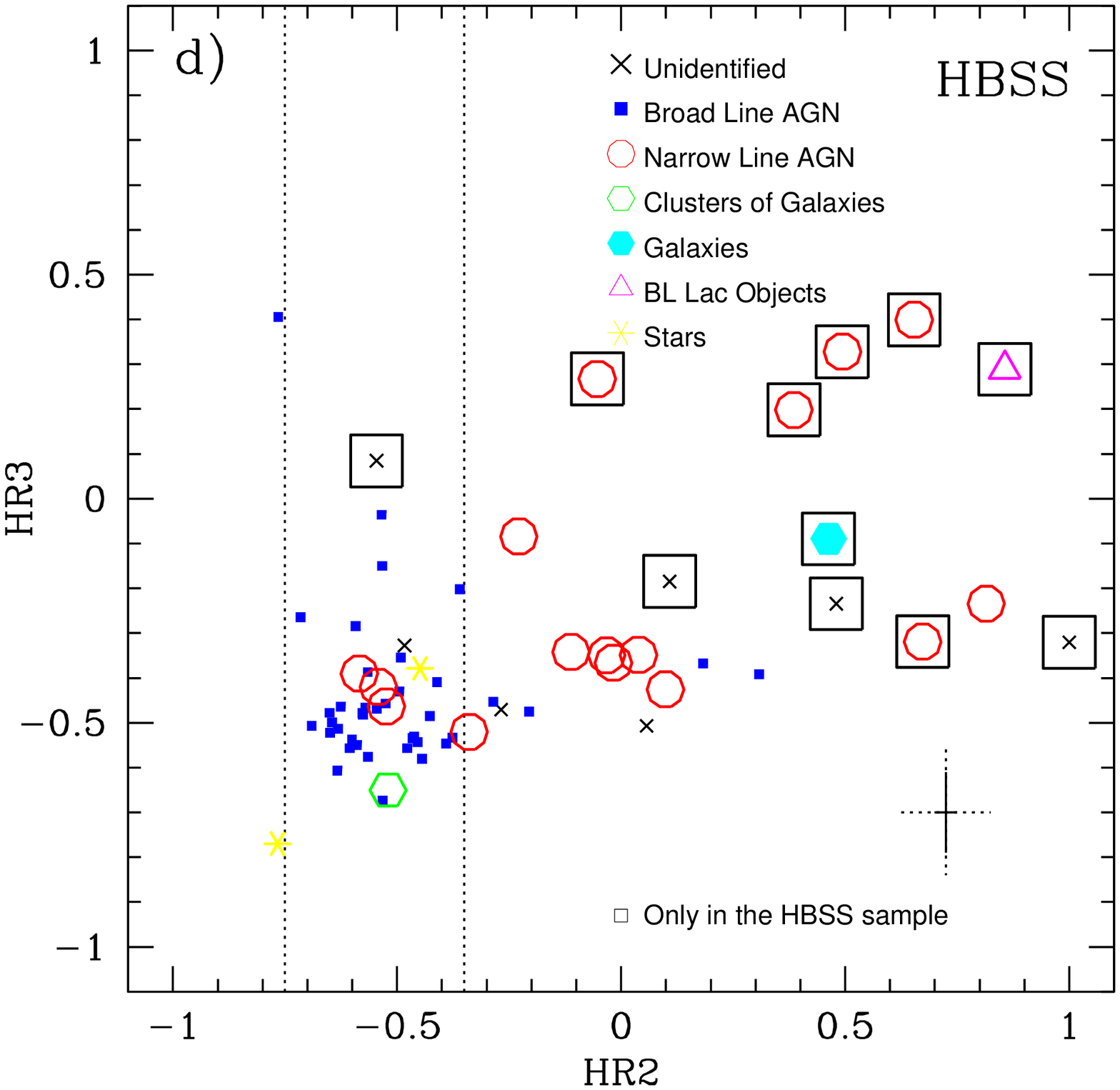}}
     \caption{
     HR2 vs. HR3 for the sources belonging to the BSS
     sample (panel a, b and c) and to the HBSS sample (panel d). The
     dotted lines at constant HR2 correspond to the locus enclosing
     $\sim 90$\% of the Type 1 AGN in the BSS sample; these lines have
     been reported in all panels to assist with the comparison(s).
     In the lower right corner of each panel we have also
     reported the median error on HR2 and HR3 for the total BSS and
     HBSS sample (solid line) and the 90\% percentile on these errors
     (dotted line). The eleven sources belonging only to the HBSS sample 
     are enclosed inside empty squares in panel d.     
               	       }  
        \label{fit1}
    \end{figure*}
%
%______________________________________________________________

\section{The X-ray to optical flux ratios}

A useful parameter to discriminate between different classes of X-ray sources
is the X-ray to optical flux ratio.  Previous investigations (e.g. Maccacaro et
al., 1988;  Fiore et al., 2003 and references therein) have shown that 
standard X-ray selected AGN (both type 1 and type 2) have an X/O flux ratio typically
in the range  between 0.1 and 10;  X/O flux ratios below 0.1 are typical of stars,
normal galaxies (both early type  and star-forming) and nearby heavly absorbed
(Compton thick) AGN; finally, at high  X/O flux ratios (above 10), we can find broad
and narrow line AGN as well as high-z high-luminosity  obscured AGN (type 2
QSOs), high-z clusters of galaxies, and extreme BL Lac objects.

In figure 2 we have plotted the X/O flux ratio versus the  HR2 value for each source
in the BSS or HBSS sample. 

The bulk of coronal emitting stars is well separated from the bulk of
extragalactic sources. Some of the AGN (both broad and narrow line)  have an X/O flux
ratio typical of stars but HR2 values typical of AGN. Similarly some AGN with
an HR2 typical of stars can be distinguished from stars thanks to their X/O
flux ratio.  Therefore the combined use of X/O flux ratio and HR2 allows us to
distinguish almost unequivocally galactic sources from the extragalactic ones, prior 
to optical spectroscopy.

Around 10\% of the extragalactic population have an X/O flux ratio $>$ 10.  If we consider the 2--10
keV fluxes instead of the  0.5--4.5 keV fluxes this fraction increases to $\sim 15$ \%, 
in good agreement with the results obtained by Fiore et al. (2003) at fainter fluxes. 
Amongst the sources with X/O flux ratio  $>$ 10 identified so far there are some broad line and a
few narrow line AGN but the large majority of the X-ray sources in this part of the X/O
diagram is  still unidentified. According to Fiore et al. (2003) some type 2 QSOs are
expected among these objects. 

The opposite side of the extragalactic X/O flux ratio distribution  (X/O $<0.1$) seems to be
populated by  optically normal galaxies, Type 2 AGNs and a few broad line AGN.

  \begin{figure*}
  \centering
  \resizebox{\hsize}{!}
    {\includegraphics[clip=true]{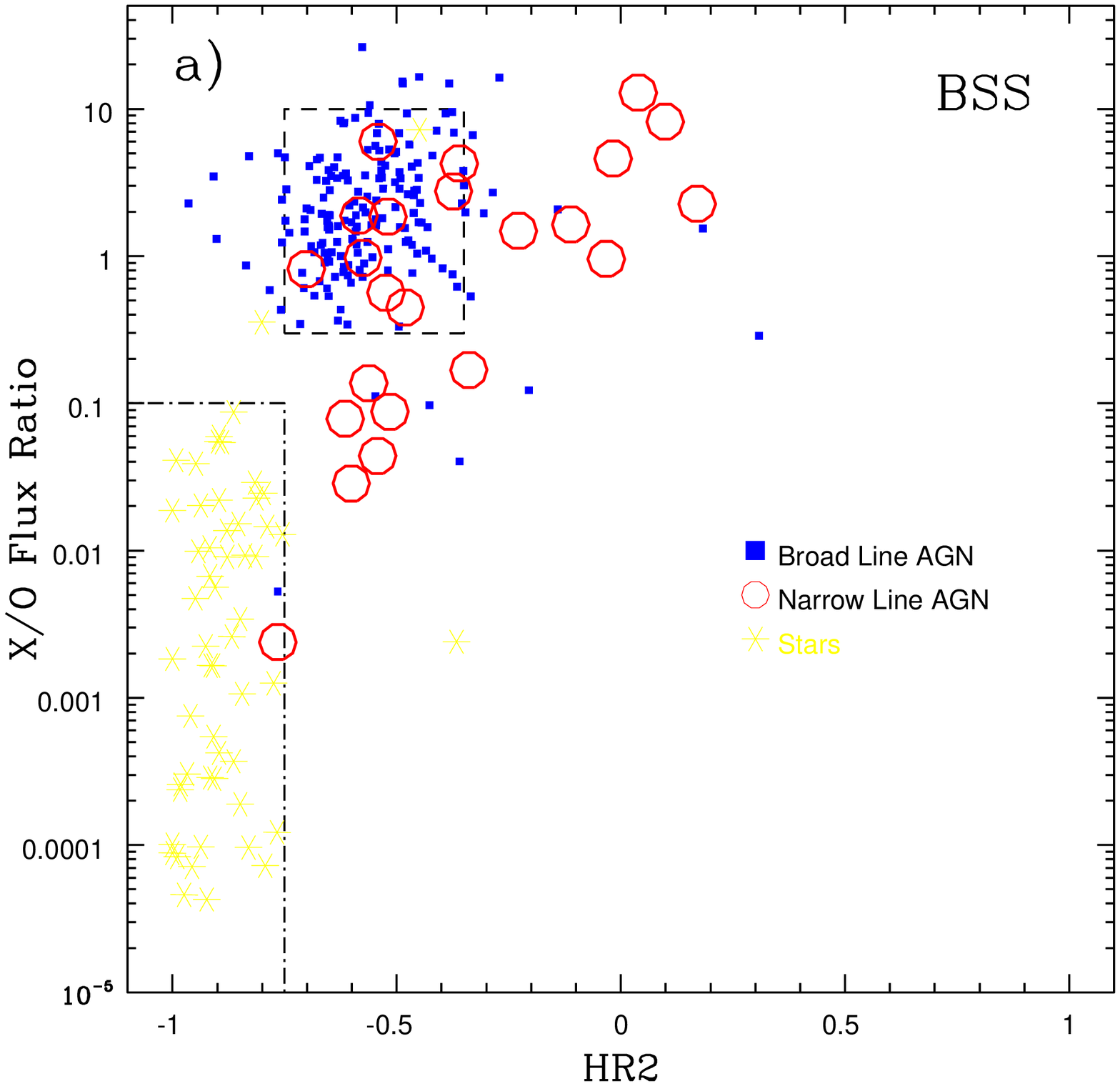}
     \includegraphics[clip=true]{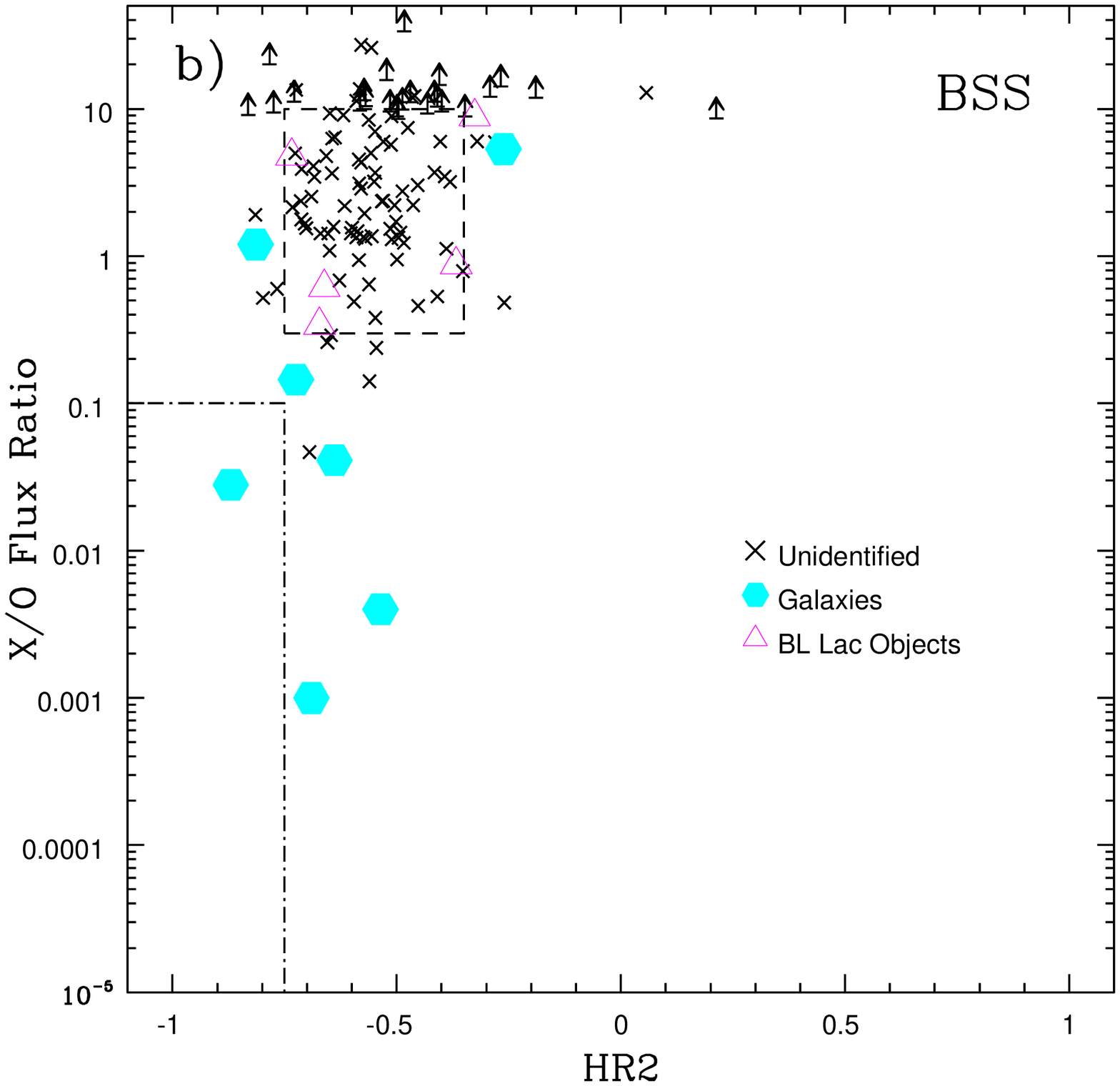}}
   \begin{tabular}{cc}
     \includegraphics[width=0.50\textwidth]{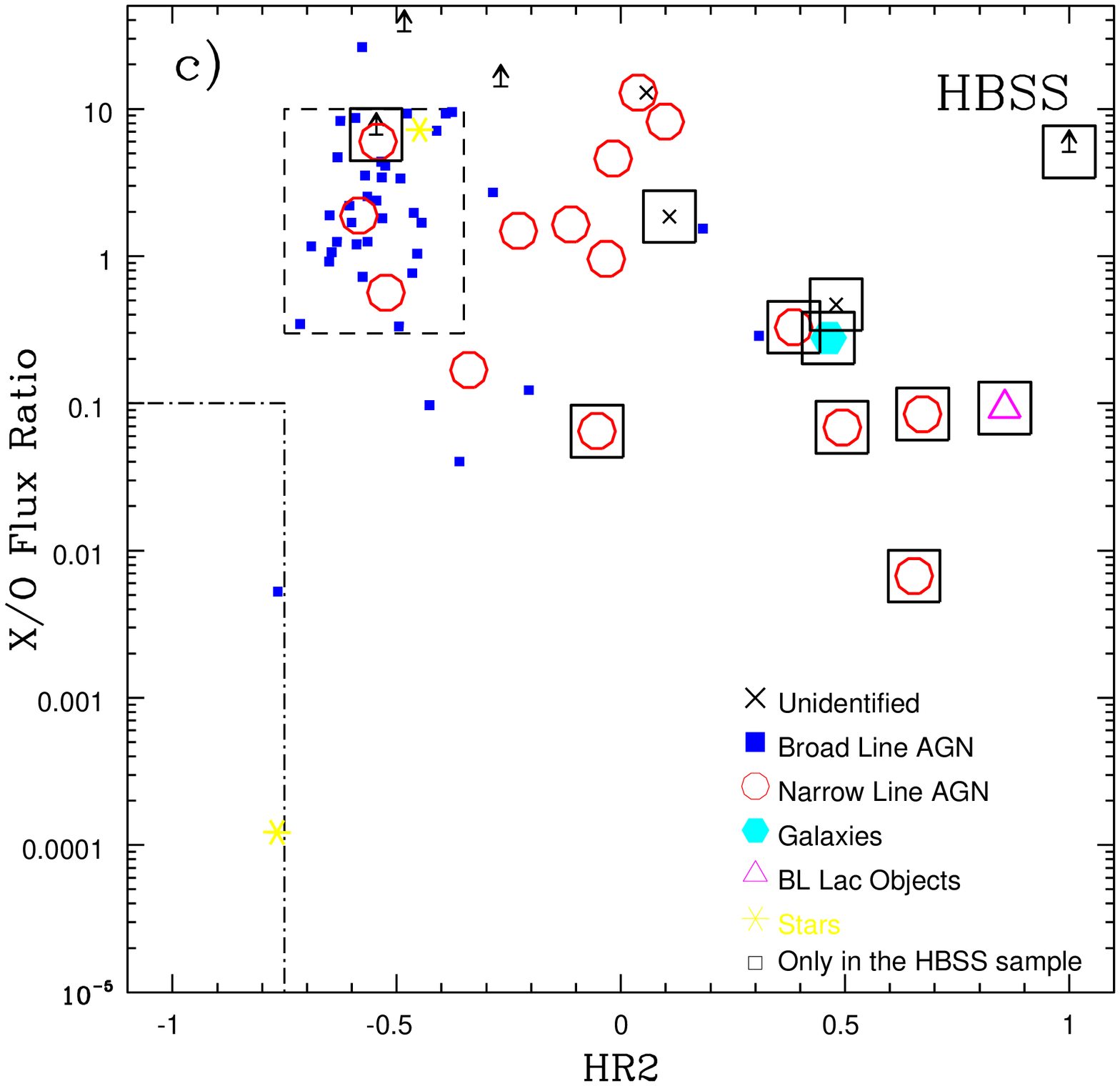}
       \begin{minipage}{7cm}
       \vspace*{-7cm}
        \caption{ X-ray (0.5-4.5 keV) to optical (APM red band)
         flux ratio vs. HR2 for the sources belonging to the BSS sample (panels
         a and b) and to the HBSS sample (panel c). The box defined by the dot-dashed
         lines (dashed line) encloses about 95\% (85\%) of the sources
         optically identified as star (broad line AGNs) in the BSS
         sample; these boxes have been reported in all panels to assist with
         the comparison(s).
         The eleven sources belonging only to the HBSS sample 
         are enclosed inside empty squares in panel c.}
      \end{minipage}\\
      \end{tabular}
      \label{fit}
    \end{figure*}

%
%                                                Two column figure
%----------------------------------------------------------- S_vib   
%   \begin{figure}
%   \centering
%   \resizebox{\hsize}{!}{\includegraphics{figura3_articolo.eps}
%   \includegraphics{figura4_articolo.eps}}
%   \caption{Left panel: Best fit of the cluster with
%            1.75 Gyr theoretical isochrone with OPAL EOS.
%            Right panel: Best fit of the cluster with 1.95 Gyr
%            theoretical isochrone with Straniero (1998) EOS} 
%     \label{fit3}
%\end{figure}
%
%______________________________________________________________

%
%                                                Two column figure
%----------------------------------------------------------- S_vib   
%   \begin{figure}
%   \centering
%   \resizebox{\hsize}{!}{\includegraphics{figura5a_articolo.eps}
%   \includegraphics{figura5b_articolo.eps}}
%   \caption{Left panel: solid line: theoretical standard (without overshoot)
%            2.0 Gyr isochrone with Straniero (1998 \cite{straniero}) EOS (Z=0.007 Y=0.244) with
%            overimposed the corresponding synthetic CM diagram including 30\% of
%            binaries; dashed line: theoretical 2.2 Gyr isochrone with Straniero
%           (1998\cite{straniero}) EOS and mild overshooting (${l}_{ov}=0.1
%           {H}_{p}$). 
%	   Right panel: NGC2420 CM diagram by Anthony-Twarog et al. 1990 \cite{anthony-twarog}.
%           Observational data are not dereddened.
%     \label{fit1}
%\end{figure}
%
%______________________________________________________________

\section{The number densities of broad and narrow line AGN}

The spectroscopic identification rate of the HBSS sample is $\sim 90$\%, allowing us to
investigate for the first time the X-ray  Log(N$>$S)-LogS of broad and narrow line AGN
in the same sample; these Log(N$>$S)-LogS have been reported in figure 3. 

At the 4.5--7.5 keV flux limit of $f_x\geq \sim 7 \times 10^{-14}$ erg cm$^{-2}$ s$^{-1}$ 
the
surface densities of type 1 AGN and type 2 AGN are $1.63\pm 0.25$ deg$^{-2}$ and
$0.83\pm 0.18$ deg$^{-2}$, respectively.   Type 2 AGN represent ($34 \pm 9$)\% of the AGN
population shining in the  4.5--7.5 keV energy band above  $f_x\geq \sim 7 \times
10^{-14}$ erg cm$^{-2}$ s$^{-1}$;  finally, at the same flux limit,  type 2 QSOs are $\sim
25$ \% of the type 2 AGN population (using a  surface density of type 2 QSOs of
$0.21^{+0.15}_{-0.13}$ deg$^{-2}$ as  measured by Caccianiga et al., 2004). 

%_____________________________________________________________
%                 A figure as large as the width of the column
%-------------------------------------------------------------
   \begin{figure}
   \centering
   \includegraphics[width=6cm]{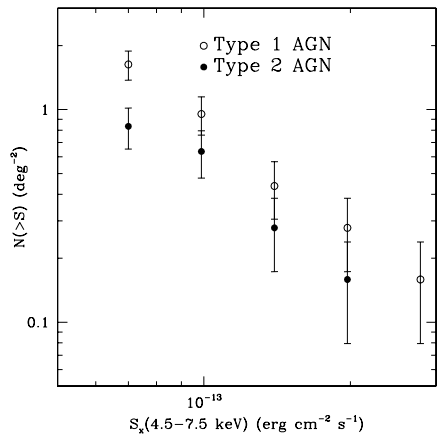}
      \caption{The number-flux relationship in the 4.5-7.5
keV energy band for type 1 and type 2 AGN.
              }
%         \label{FigVibStab}
   \end{figure}
%
%_____________________________________________________________

\section{Conclusions}

We have presented results of a survey project started few years ago. 
Full details of the results presented here, 
along with the basic information on the complete sample of 400 sources,
can be found in Della Ceca et al. (2004). 
Other results have been also discussed in Severgnini et al. (2003)
and Caccianiga et al. (2004).

Since the investigated sample is a fair representation of the high galactic latitude
X-ray sky, the results from this  project will help with the candidates selection 
of interesting 
classes of sources from the XMM-Newton catalogue prior to spectroscopic observations, 
making the existing and incoming XMM-Newton catalogs
\footnote{see http://xmmssc-www.star.le.ac.uk}
an unique resource for
astrophysical studies.

\begin{acknowledgements}
RDC, TM, AC, PS, VB acknowledge partial financial support by the Italian Space
Agency (ASI grants: I/R/037/01, I/R/062/02 and I/R/071/02), by the MURST
(Cofin-03-02-23) and by INAF. We thank the TNG Time Allocation Committee for a generous and
continuos  allocation of observing time.
\end{acknowledgements}

\bibliographystyle{aa}

\end{document}